\documentclass[reprint,twocolumn,letterpaper]{revtex4} 

\usepackage[ruled]{algorithm2e}
\usepackage{graphicx}
\usepackage{subfigure}  

\usepackage[normalem]{ulem}

\SetAlFnt{\small}
\SetAlCapFnt{\small}
\SetAlCapNameFnt{\small}
\SetAlCapHSkip{0pt}
\IncMargin{-\parindent}

\begin{document}

\date{13 November 2015}

\title{High-Performance Computing with Quantum Processing Units}
\author{KEITH A. BRITT}
\affiliation{University of Tennessee and Oak Ridge National Laboratory}
\email{keithbritt@utk.edu}
\author{TRAVIS S. HUMBLE}
\affiliation{Oak Ridge National Laboratory and University of Tennessee}
\email{humblets.ornl.gov}

\thanks{This manuscript has been authored by UT-Battelle, LLC, under Contract No. DE-AC0500OR22725 with the U.S. Department of Energy. The United States Government retains and the publisher, by accepting the article for publication, acknowledges that the United States Government retains a non-exclusive, paid-up, irrevocable, world-wide license to publish or reproduce the published form of this manuscript, or allow others to do so, for the United States Government purposes. The Department of Energy will provide public access to these results of federally sponsored research in accordance with the DOE Public Access Plan.}

\begin{abstract}
The prospects of quantum computing have driven efforts to realize fully functional quantum processing units (QPUs). Recent success in developing proof-of-principle QPUs has prompted the question of how to integrate these emerging processors into modern high-performance computing (HPC) systems. We examine how QPUs can be integrated into current and future HPC system architectures by accounting for functional and physical design requirements. We identify two integration pathways that are differentiated by infrastructure constraints on the QPU and the use cases expected for the HPC system. This includes a tight integration that assumes infrastructure bottlenecks can be overcome as well as a loose integration that assumes they cannot. We find that the performance of both approaches is likely to depend on the quantum interconnect that serves to entangle multiple QPUs. We also identify several  challenges in assessing QPU performance for HPC, and we consider new metrics that capture the interplay between system architecture and the quantum parallelism underlying computational performance.
\end{abstract}

\maketitle

\section{Introduction}
\label{sec:intro}
High-performance computing (HPC) has historically taken advantage of new processing paradigms by leveraging special purpose accelerators. This includes the use of algorithmic logic units and floating-point units in early processor architectures as well as more recent vector processors capable of single-instruction multiple-data (SIMD) parallelization. Current interest in graphical processing units (GPUs) is another example of the ongoing trend in accelerator use for HPC development. A primary motivation for the accelerator paradigm is that low-level processes can take advantage of specialized hardware while minimizing changes to overall program structure \cite{Kindratenko2008}. This approach isolates the need for program or algorithm refactoring to those workloads specific to the hardware accelerator \cite{Schneider2015}. A secondary motivation is that the accelerator model offers an opportunity to take advantage of emerging technologies while also mitigating the technical risk to system development. Current limitations on processor frequency, communication bandwidth, physical scale, energy consumption, and hardware reliability make it advantageous for HPC designers to anticipate new technologies and to leverage architectures that support innovative platforms. Future efforts to realize HPC beyond the current exascale target are likely to require such innovations  \cite{Lucas2014,Ashby2010,Geist2009}.
\par
The search for technology paths that lead to performance beyond exascale may require alternative computational models. This is because some problems are better suited to computational models other than the standard Turing machine model. In particular, quantum computing has attracted significant interest due to theoretical results that exponential reductions in algorithmic complexity of some problems are possible relative to the best known conventional algorithms \cite{Simon1997}. This includes the factorization of integers, a staple of public-key cryptography \cite{Shor1997}; ab initio calculations of electronic structure in chemistry and physics \cite{Kassal2010}; and scattering amplitudes of particles in high-energy physics \cite{Abrams1997}. These algorithmic speedups are achieved by leveraging the unique features of quantum mechanics, namely: superposition, entanglement, and intrinsic randomness. The basic principles of quantum computing have been demonstrated in small-scale experimental systems and there is an on-going, global effort to develop large-scale quantum computing platforms.
\par
The opportunities afforded by quantum computing represent a challenge to the HPC accelerator model, which previously has been practiced exclusively within the setting of the classical, deterministic Turing model. By contrast, many quantum algorithms lack clearly defined kernels that can be off-loaded to a quantum computational accelerator. The mixture of computational models also stymies efforts to leverage conventional notions of SIMD and multiple-instruction multiple-data (MIMD) parallelism. Parallel computing typically makes use of domain decomposition \cite{vanNieuwpoort2001} whereas quantum algorithms frequently intentionally avoid this type of problem partitioning \cite{Bauer2015,Kreula2015}. In addition, domain decomposition exposes an interface between classical and quantum computational models that is not yet well defined. Translation between computational models may be theoretically possible but making these interfaces efficient and robust is an outstanding concern.
\par
As constraints on the physics underlying quantum computation limit how these resources may be used, it is unclear if and how emerging quantum processors will become compatible with existing or future HPC platforms. We analyze the integration of these quantum processing units (QPUs) for modern HPC architectures. We place an emphasis on conceptual differences between the conventional and quantum computing models that may be expected to challenge integration. Our analysis examines two pathways that lead to several abstract machine architectures. We identify those distinguishing features that QPUs may be expected to exhibit and the dimensions that will be most useful for characterizing their performance metrics within a hybrid system.
\par
The paper is organized as follows. In Sec.~\ref{sec:qpu}, we characterize the features of a QPU, briefly summarize its operating principles, and identify requirements for operation that influence HPC integration. In Sec.~\ref{sec:qpuip}, we examine three multi-processing models for adopting QPUs into conventional HPC systems and the architectures that arise from them. In Sec.~\ref{sec:metrics}, we describe the need for both standardized as well as unique performance metrics to characterize HPC with quantum accelerators. We offer final remarks in Sec.~\ref{sec:conclude}.
\section{Quantum Processing Unit (QPU)}
\label{sec:qpu}
\par
We define a quantum processing unit (QPU) to be a computational unit that uses quantum computing principles to perform a task. As the operating principles of a QPU are based on quantum mechanics, there are several unique features that do not have analogs in conventional computing platforms. Foremost, a QPU stores computational states in the form of a quantum mechanical state. While a quantum state is formally defined as a unit vector in a finite-dimensional Hilbert space \cite{Nielsen2000}, it must also be interpreted as the data processed by the QPU. The simplest and most frequently used example is a qubit, which expresses a state within a two-dimensional Hilbert space. These states are stored in quantum physical systems. For the purpose of clarity, we define a quantum register as an addressable array of two-level quantum physical systems. We will refer to an individual system within the register as a quantum register element and we will assume that each register element can store a qubit of information. We may refer to the size of the register by the number of qubits that it can store, e.g., an $n$-qubit register.
\par
The computational space available to a quantum register scales exponentially with its size. Like an $n$-bit register, an $n$-qubit register is capable of representing all $2^n$ computational states. However, the quantum register is also capable of representing superpositions of these states, a phenomenon known as entanglement. Fundamentally, entanglement is a limitation on the ability to describe states of a register solely by specifying the value of each register element. This is in stark contrast to classical models of computation and leads to a description of what has been called the `inherent parallelism' of quantum computing \cite{Deutsch1985}. This inseparability of register states manifests as perfect correlations during computation, and many quantum algorithms take advantage of entanglement to realize computational speed ups \cite{Childs2010}.  
\par
Operations on the quantum register are realized using gates. Like conventional computing, quantum gates correspond with well-defined transformations of the computational state. When the register is prepared in a superposition state, operations effectively act on multiple computational states in parallel. This may be viewed as a quantum variant of conventional SIMD processing. However, quantum computing makes use of gates that are either unitary transforms of the register elements or projective measurements. Only the latter gates prepare the state of the quantum register in a well-defined classical value, e.g., either a 0 or 1 for each register element. For a unitary gate, the value of the register remains in a superposition of computational states and serves as an intermediate computation. Ultimately, the solution to a computation is recovered when a projective measurement gate is applied to the register. The resulting bit string must then be stored in a classical register within the QPU.
\par
Quantum computational models define how the registers and gates within a QPU realize quantum computation. While all the models offer identical computational power from a complexity perspective, they do differ with respect to hardware implementations and principles of operation. For example, the quantum circuit model is closely related to the conventional representations of classical circuits, as it uses sequences of discrete gates acting on registers to generate a series of computational states. By contrast, the adiabatic quantum computing model uses a continuous, time-dependent transformation of the interactions between register elements to evolve the computational state toward a solution. For example, recent special-purpose processors within  the adiabatic quantum computing model implement quantum optimization using a single instruction that is tunable in duration \cite{Johnson2011}. Across all computational models, QPUs require precise control over the quantum physical degrees of freedom defining register elements. There is an ongoing effort to demonstrate proof-of-principle registers and gates within a variety of physical systems, including silicon donor systems \cite{Saeedi2013}, trapped ions \cite{Monroe2013}, and superconducting circuits \cite{Devoret2013}. A specific focus has been on realizing high-fidelity implementations that can support fault-tolerant operation. In addition, there has  been some work to design the physical layout and instruction architectures for certain technologies \cite{vanMeter2006,Thaker2006,vanMeter2010,Jones2012,Metodi2011,Hill2015}. 
\par
A typical usage case for a QPU begins by preparing the quantum register in a well-defined initial state and then applying a sequence of gates that may act on individual or multiple register elements. This is commonly referred to as the QRAM model, first articulated by Knill \cite{Knill1996}. We define a QPU to include a QRAM that applies low-level gates to register elements, a quantum control unit (QCU) that parses programs into instructions, and a classical controller interface that defines how the CPU within a host system interacts with the QPU. The exact sequence of gates is determined by the host program, which the QCU parses into an intermediate representation using an instruction set architecture (ISA). The ISA represents a set of high-level instructions that are available for programming the QRAM. Instruction sequences are generated when a compiled program is decoded by the QCU, after which the instructions are parsed by the QRAM into gates, i.e., machine codes, that are specific to the QPU technology base. At present, there are a variety of technologies under consideration for quantum computing and they each support different gate operations. The ISA provides a framework for standardizing the interface between different QPU technologies.

\begin{figure}[ht]
\centerline{\includegraphics[width=8cm]{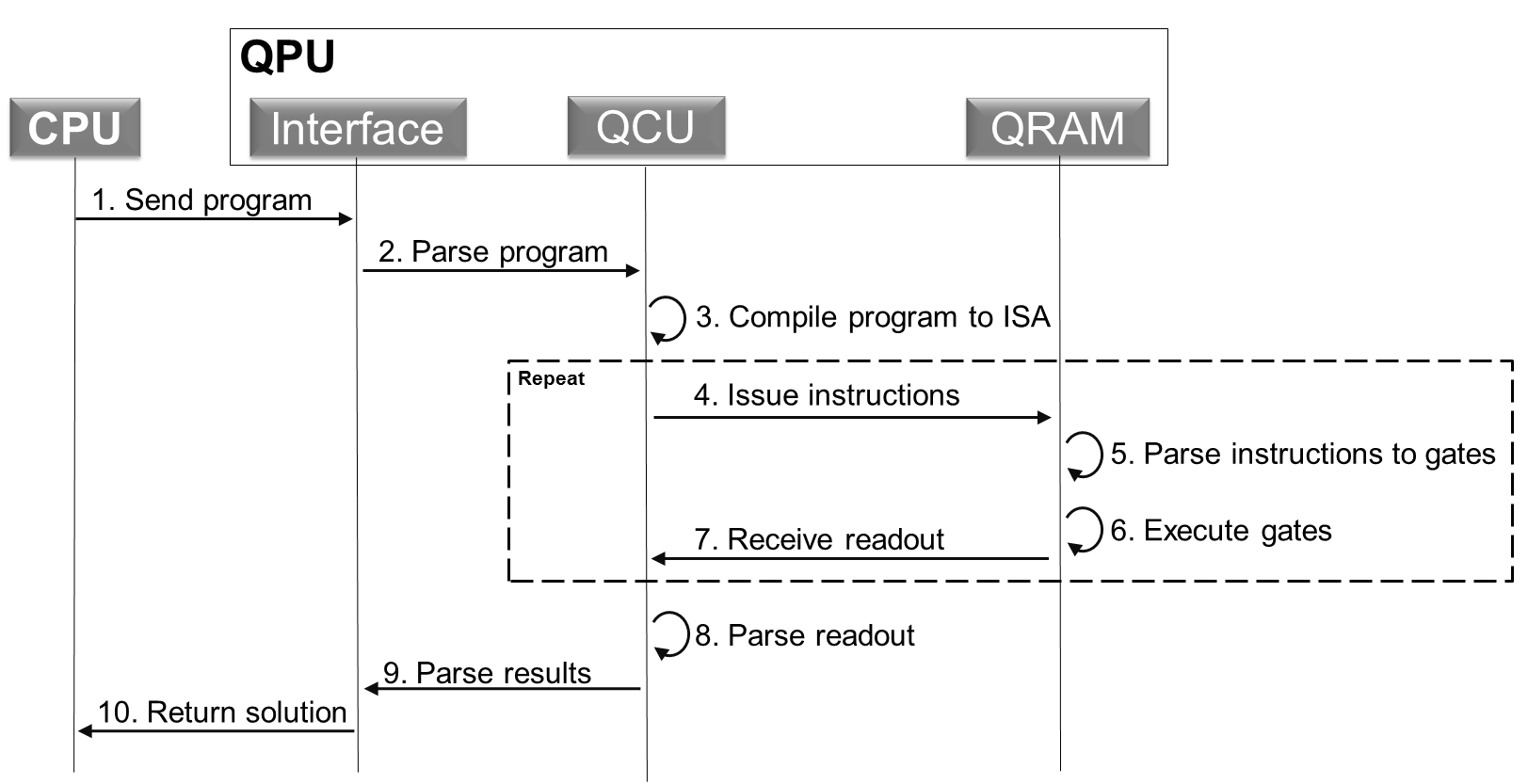}}
\caption{A sequence diagram modeling the interactions between a host CPU and a QPU. The QPU interface defines the internal QRAM model and drives gates operating on the register. The QCU parses the incoming instructions and the outgoing response according to the computational model and device physics.}
\label{fig:qram}
\end{figure}

\par
Applying the QRAM model within the QPU context defines an interface with the classical controller allocated to the host CPU (see Fig.~\ref{fig:qram}). The CPU tasks the QPU by submitting a program and then waits for the reply. These tasks must represent quantum computational workloads that can be parsed out by the QCU, where the interface may be implemented in software or hardware. Development of these interfaces is still an open question although recent progress has been made in defining quantum programming languages for this purpose \cite{Selinger2004,Green2013,Wecker2014,Abhari2012}. These languages offer exposure to the gates and registers needed for programming low-level quantum algorithms, and we expect future generations will likely grow to addressing additional data structures and instructions. Interfaces that encapsulate and mask quantum hardware details from the software developer are especially important for maintaining existing applications across a variety of HPC environments. As an example, an application program interface (API) that requires software developers to integrate quantum processing instructions directly into the code base will prevent its adoption (due to the burden of code rewriting).
\par
In addition to local operations driven by a host CPU, a QPU may also interact directly with other QPUs. This may be necessary to communicate a computational state between QPUs or to prepare both registers in a mutually entangled state. These operations require the presence of a quantum network, which uses quantum physical systems to communicate quantum states between registers. However, a notable feature of quantum computing is that intermediate computations cannot be copied during the communication. This is a consequence of the no-cloning theorem from quantum mechanics that limits the precision with which arbitrary quantum states can be duplicated. Instead, communication between QPUs must use either direct transmission or teleportation \cite{Nielsen2000}. Direct transmission transfers the value of the first register over the quantum network by using a mobile quantum carrier. Upon receiving the transmission, the second QPU swaps this states into its register. This approach most closely resembles conventional read-write communication in an HPC network. But quantum communication also supports teleportation, which allows for the value of a first register to be transferred to a second register without passing through the quantum network. Instead, teleportation uses pre-existing entanglement between the QPUs to perform the data transfer. This does also require transmission of classical side-channel information from the first QPU to the second, however this classical information is generally much less than the information needed to describe the value of the quantum register. 
\section{QPU Integration Strategies}
\label{sec:qpuip}
\begin{figure}
\centerline{\includegraphics[width=8cm]{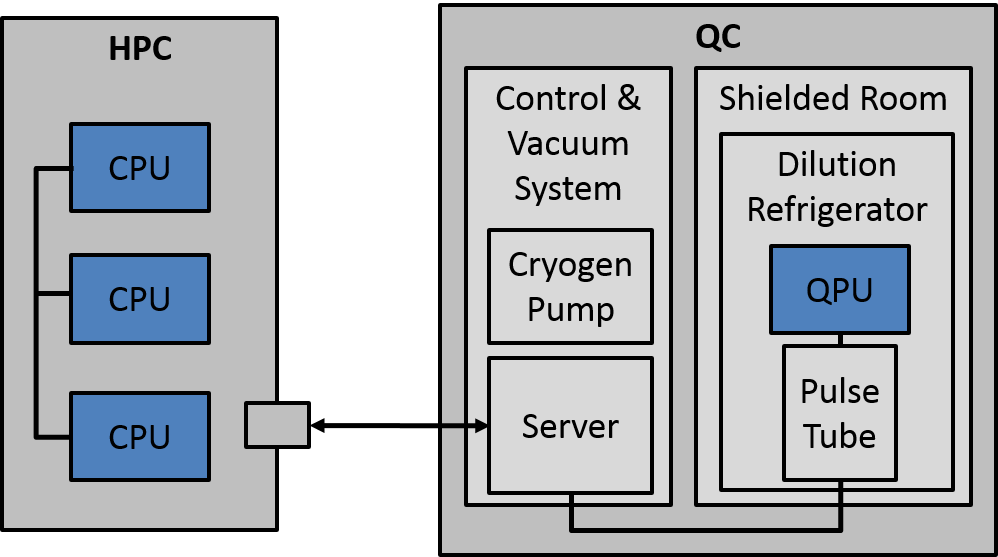}}
\caption{Asymmetric multi-processor architecture for integrating a stand-alone quantum computer (QC) with an HPC system. We highlight components of the QC system that represent the substantial infrastructure required for interfacing and controlling the QPU.}
\label{fig:asymulti}
\end{figure}

The simplified CPU-QPU execution model presented in Sec.~\ref{sec:qpu} offers a variety of different integration strategies for the development of large-scale hybrid computing systems. A significant obstacle to integration is the physical hardware requirements of current experimental quantum computing devices. Many technologies that could be used to realize a fully functional QPU currently require bulky and costly infrastructure. This includes the use of dilution refrigerators to suppress thermal noise, electromagnetic shielding to avert ambient energy, and ultra-high vacuum enclosures to prevent device contamination. In addition, most devices require relatively complex electronic and optical control systems that must cross the physical barriers to the processor. An exemplary system schematic is shown in Fig.~\ref{fig:asymulti}.  
\par
We anticipate that QPU requirements will ease with future device development and refined engineering principles. For example, recent work on ultra-cold operation of FGPAs to drive silicon qubits suggest there is a path toward integration of the QPU control interface within the dilution refrigerator \cite{Hornibrook2015}. Similarly, progress in the miniaturization of electronic controls for linear optical quantum chips hints at scalable operations in the future \cite{Carolan2015}. However, these devices still remain far from the typical hardware environments on which modern HPC systems have been built, namely, room temperature operation, direct interaction with the host CPU, and easily managed footprints for individual processors. Consequently, integration opportunities for QPUs naturally separate into loosely and tightly bound systems. 
\par
\begin{figure}
\centerline{\includegraphics[width=8cm]{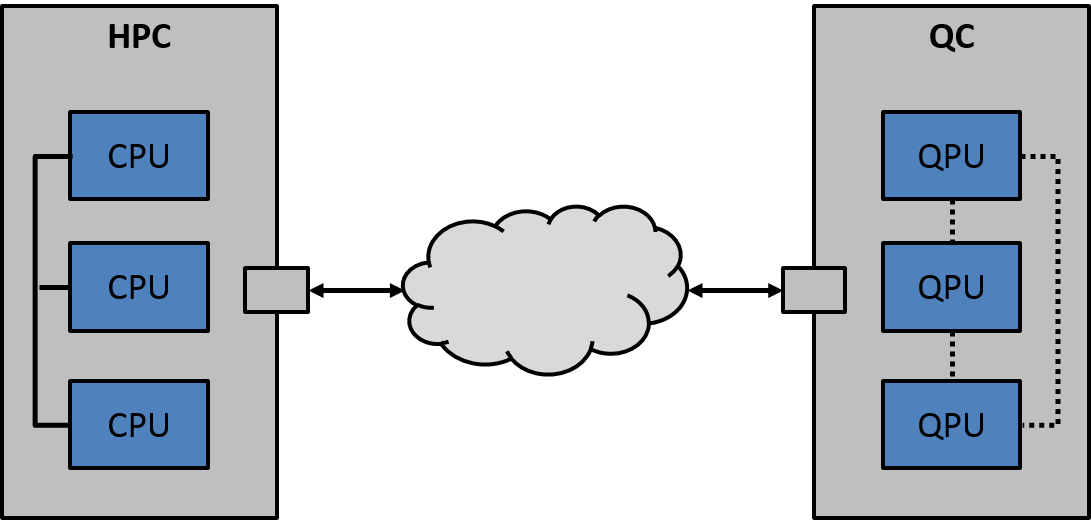}}
\caption{An asymmetric multiprocessor model employing a quantum computing (QC) server, e.g., a form of cloud-based quantum computing. The dashed lines indicate a quantum interconnects between QPUs while solid lines indicate classical interconnects. The concept of QC as a service offers increased flexibility and ease-of-use at the expense of communication latencies. Latencies will contribute to overall execution timing and, depending on problem and program structure, could partially negate quantum computational advantages.}
\label{webservice}
\end{figure}

In the loose integration path, QPUs remain as isolated operational elements that must interact with a host HPC system using a network interface. This is effectively a client-server model as shown in Fig.~\ref{webservice} where the quantum computing (QC) server may either be on a dedicated network or part of a larger computational grid. In this asymmetric multi-processor model, the network communicates requests between the host (client) system and the QC server. As indicated in Fig.~\ref{webservice}, the QC server may host multiple QPUs and these may interact via a quantum interconnect. However, the entry point into the system remains the primary bottleneck. This connection can be streamlined when both systems are within a local area network, but access to each QPU must still be provisioned by the QC control system. This control system may appear as a switch that forwards program data to individual QPUs, or it may more intelligently route programs based on QPU usage and demand.
\par
The demands of the client-server model in Fig.~\ref{webservice} force a trade-off between the communication latency and the computational speed-up gained from using a QPU instead of a CPU (or some other local resource). This trade-off is advantageous when the communication time is offset by the QPU speed up, but this will depend on problem type as well as size. Moreover, evaluation of the model is complicated by the communication patterns arising from multiple CPU nodes and any latency they may experience from resource competition. Therefore, the client-server model is likely to be broadly useful only when the computational gain over conventional approaches is significant. This adds emphasis to the importance of the underlying quantum algorithm.
\par
The loosely integrated client-server model also supports the alternative use case of cloud-based quantum computing. In this setting, the QC server is a rare resource in demand from multiple users simultaneously. For a system containing $q$ QPUs, the server can support classical MIMD parallelism with each node performing an isolated job. As a measure of server capacity, the dimension of the server Hilbert space is $q 2^n$ given a $n$-qubit register on each node. By contrast, the presence of a quantum interconnect linking the individual QPUs offers a Hilbert space of $2^{n q}$. This exponential increase in server capacity with node number is not a guarantee of computational speed up. The cloud-based quantum computing model may be especially attractive for blind quantum computing, which permits a user to submit a job request without revealing details about either the data or instructions \cite{Broadbent2009}.
\par
\begin{figure}
\subfigure{
\includegraphics[width=4cm]{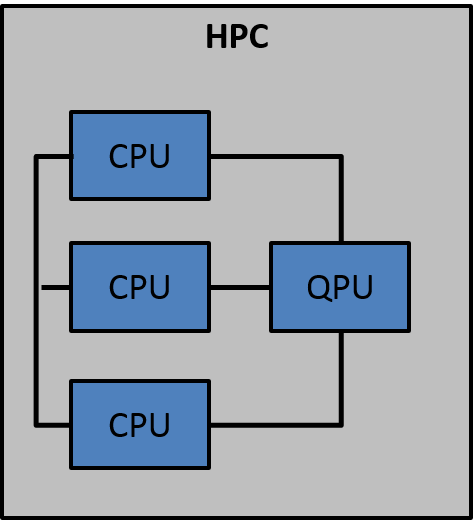}}
\subfigure{
\includegraphics[width=4cm]{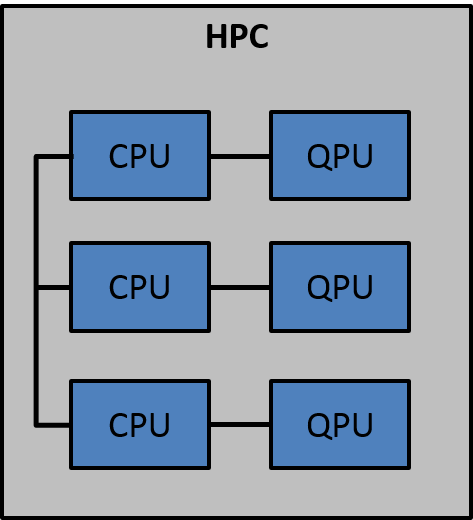}}
\caption{(left) A shared resource model in which a single QPU is accessed by multiple CPU nodes. (right) A standard accelerator model in which QPUs are attached to nodes hosted on a classical interconnect. The absence of quantum networking between QPUs restricts the scaling with respect to the quantum resources and enforces a classical domain decomposition paradigm. }
\label{Figure_04}
\end{figure}
\begin{figure}
\centerline{\includegraphics[width=4cm]{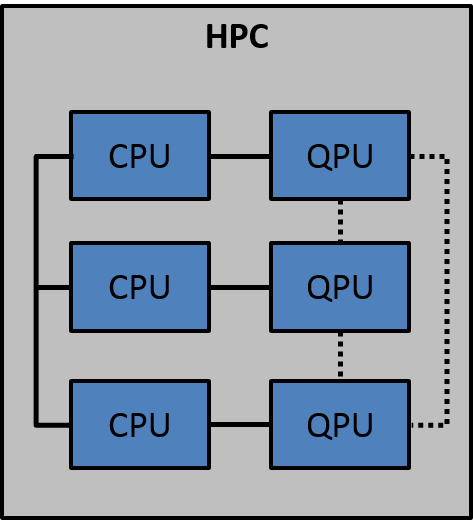}}
\caption{An accelerator model with QPUs incorporating a quantum interconnect that supports both quantum and classical parallelism. QPUs may be addressed individually or collectively through the coordinated CPU elements.}
\label{final_diagram}
\end{figure}

The tight integration path is shown in Figs.~\ref{Figure_04} and \ref{final_diagram} and represents a progression toward more sophisticated accelerator models.  The goal of this design is to move the QPU closer to the host node in order to eliminate communication latency  and maximize computational speed up. This design assumes the hardware requirements for QPUs can be eased to the point that a single, tightly connected single-system model can be created. As mentioned above, this will require multiple advances in the classical infrastructure used by current experimental devices. Figure~\ref{Figure_04} (left) represents a first design based on a shared resource model that permits multiple CPU nodes to interact with a single QPU node. Like the server model, a single QPU is responsible for managing  requests for multiple CPUs and requires a robust classical controller interface. This communication design also represents a bottleneck but the tighter integration alleviates some of the overhead required in the loose model. In addition, data from multiple CPUs can be aggregated by the QPU. This use case may appear when pre-processing of the input for the QPU can be parallelized across CPU nodes. Alternatively, if the QPU is part of a MIMD or data streaming model, then the redirection of QPU results to another node may be useful.

\par
A more pronounced example of the accelerator model is presented in the right panel of Fig.~\ref{Figure_04}. This design most closely matches that used for integrating GPU accelerators into modern HPC systems as each CPU node is tightly integrated with a dedicated QPU. This greatly simplifies the QPU interface. This hardware model also naturally matches many existing program data access patterns, in which top-level memory management is driven by domain decomposition with low-level data movement restricted to single nodes. In this sense, the design is motivated by an initial application of classical parallelism and subsequently followed by quantum parallelism. While this model offers the appeal that it would minimize the refactoring required of existing source codes, it also restricts the amount of quantum parallelism available.
\par
Instead, Fig.~\ref{final_diagram} represents an accelerator model in which individual QPUs are interconnected. The quantum interconnect establishes quantum communication between each QPU and offers the possibility of generating entangled states between registers. Like the conventional interconnect used to establish communication between CPU nodes, the quantum interconnect will require additional switches and possibly routers to perform robust communication. In this tightly integrated limit, a collection of interconnected QPUs may be abstracted as a single QRAM accessed through interfaces at multiple CPU nodes. This design offers the benefit of maximizing the potential quantum parallelism across QPUs and the classical parallelism across CPUs, where hybrid nodes may cooperate by exchanging messages and data across both interconnects.
\section{Performance Metrics}
\label{sec:metrics}
An important aspect in evaluating the merits of QPUs for use in any of the system integration strategies is identifying performance metrics that are both well-defined and meaningful. Basic metrics for conventional computing typically focus on register size, word size, FLOPS, etc. More elaborate measures take account of multi-threaded processor pipelines, memory access speeds, and communication latency. Processing models that include SIMD and MIMD parallelism are also judged according to use cases and targeted problem sets. Each metric can also be monetized, e.g., FLOPS per watt and FLOPS per unit currency, to address specific  stakeholder interests.
\par
Despite a long history of evaluating conventional HPC systems, the unique features of QPUs pose new challenges in measuring performance. First and foremost is the difference in the underlying computational model. For example, QCUs and QRAMs will certainly require clocks to execute the instructions and machine codes acting on the quantum register, but the speed of the clock is not directly proportional to the speed at which gates are applied. Instead, the gates themselves are complicated sequences of control signals applied to the physical system and they cannot be made arbitrarily fast. The speed of gate execution is complicated further by the use of fault-tolerant (FT) instruction protocols. These protocols protect against errors but necessarily require the use of additional primitive gates and qubits \cite{Nielsen2000}. Fault-tolerant instruction protocols may vary with program sequence as well as data locality. Since the context of the program ultimately determines the speed at which gates are executed, the quantum FLOPS analogy for measuring QPU performance is poorly defined.
\par
Nevertheless, QPU performance can be measured. One measure is the relative overhead required by the FT instruction protocols for the ISA. At the scope of the QCU, the cycles per instruction can be extracted. Moreover, the longest duration gate within a QRAM can serve as a worst-case measure of performance while the shortest gate reflects best time cost. The timing difference between these instructions offers a measure of the spread in QRAM performance on any random gate instance. The same measures can be applied to the complete set of QPU instructions to compile a snapshot of worse and best-case timings.  
\par
These definitions do not require reference to a particular technology as they rely on abstraction of the QRAM and QPU interfaces. This is advantageous for making performance comparisons against different technologies, which may have widely different physics and control signals. Different QPU technologies may also employ vastly different ISAs in a manner that is reminiscent of RISC versus CISC designs. For example, a QPU based on the adiabatic quantum computing model may use a single, time-continuous gate to implement an instruction, whereas the same instruction would be realized in the circuit model using a lengthy sequence of discrete primitive gates. A comparison between these different QPUs that simply references gate duration and spread forgoes these important details. Additionally, task-specific QPUs will necessitate special-purpose metrics to differentiate between how problems are solved. Restrictions on processor behavior can be useful, however, for purposes of forming comparisons provided that the context is well specified.
\par
A third consideration for measuring QPU performance is that quantum algorithms are often probabilistic. This introduces the notion of repeated sampling of the readout register in order to collect sufficient statistics for reporting a final result. Depending on the level of instruction abstraction, the programmer may handle this type of sampling or it may arise within the QPU itself. For those use cases where quantum behavior is intended to be hidden from the user, e.g., with high-level languages and libraries, QPU performance will be impacted by these classical pre- and post-processing steps. The use of statistical sampling to derive the final result requires confidence levels that determine the number of required samples and therefore total duration of the program. Neither the instruction or gate metric that we have proposed would measure this aspect of QPU performance. Instead this becomes an element of benchmarking against certain problem sets and solution goals. Algorithmic complexity statements will offer some guidelines on the scaling of these slowdowns, but experimental tests will be needed to identify the variance in total function performance.
\par
Performance of the quantum interconnect is also a major factor in overall system behavior. The rate at which QPUs establish entangled registers may be initially ignored, since these operations can occur offline from the program execution. However, in a high-performance setting the rate of entangling and preserving entanglement between nodes represents a potential bottleneck for the system. Idle registers in a QPU require active error correction to mitigate against noise, and any delays in the quantum interconnect will add to this overhead. 
\par
It therefore seems very likely that QPUs will need to use some form of network management controller that interacts with the quantum programs wishing to execute using entangled registers. The network manager will be responsible for ensuring availability of entangled registers when requested by a program. This will require coordination between the interconnect, the error-corrected registers, and the program instructions, such that the manager refreshes entanglement between QPUs only when needed by the program. However, faults and latencies in the both the QPUs and the interconnect will complicate these instructions and may eventually lead to communication failure. Therefore, the performance of the interconnect and especially the entangling operations is likely to be a major factor in overall system behavior.
\section{Conclusions}
\label{sec:conclude}
As QPUs mature from basic scientific testbeds to robust devices, they will likely be adopted for both application-specific and general purpose computing. We have investigated several possible abstract architectures for integrating QPUs into HPC systems. We have examined both loose and tight integration designs, which differ primarily in the communication infrastructure and run-time environment needed to host the QPU. The relative performance of each design is expected to depend on how well the quantum algorithm and its programming model offsets the costs of this communication as well as the intended use case.
\par
We have also emphasized that one of the most important aspects of future HPC with QPUs is the quantum interconnect. It has long been appreciated for massively parallel processing systems that the communication backbone between nodes plays a significant role in performance. This has been underscored in recent years with awareness that communication costs may often be bottlenecks in application scaling. It is clear that a quantum interconnect can enhance system functionality by enlarging the set of accessible register states, but it remains unclear if the interconnect would provide a net benefit. This is because the entanglement established between nodes by the interconnect would expose the system to a potentially more serious fault model, in which correlated errors lead to a cascade of failures across QPUs. Appreciable attention has been placed on fault-tolerant ISAs within the context of local computation, and similar techniques for managing distributed QPUs will be an important issue moving forward. Fault models for these architectures and protocols for mitigating against these types of failures are needed.
\par
Existing metrics for conventional computing do not capture all aspects of the QPU behavior, and we have suggested several new features that need to be tracked when tuning system performance. This includes the overhead in fault-tolerant ISA's as well as the spread in instruction timings. However, some instructions may be so complex that there performance can only be measured in very restrictive settings, e.g., as special-purpose QPUs. The comparison of these metrics with each other offers a quantitative means of assessing the value of QPU-enabled systems, but only when they can be related to existing system metrics, e.g., FLOPS, etc. Putting quantum metrics at the same level of inspections as those of a CPU-based measurements will require more detailed execution models for the entire system.

\bibliographystyle{aipauth4-1}
\bibliography{ARXIV_Manuscript}

\begin{thebibliography}{33}
\expandafter\ifx\csname natexlab\endcsname\relax\def\natexlab#1{#1}\fi
\expandafter\ifx\csname bibnamefont\endcsname\relax
  \def\bibnamefont#1{#1}\fi
\expandafter\ifx\csname bibfnamefont\endcsname\relax
  \def\bibfnamefont#1{#1}\fi
\expandafter\ifx\csname citenamefont\endcsname\relax
  \def\citenamefont#1{#1}\fi
\expandafter\ifx\csname url\endcsname\relax
  \def\url#1{\texttt{#1}}\fi
\expandafter\ifx\csname urlprefix\endcsname\relax\def\urlprefix{URL }\fi
\providecommand{\bibinfo}[2]{#2}
\providecommand{\eprint}[2][]{\url{#2}}

\bibitem[{\citenamefont{Kindratenko et~al.}(2008)\citenamefont{Kindratenko,
  Thiruvathukal, and Gottlieb}}]{Kindratenko2008}
\bibinfo{author}{\bibfnamefont{V.}~\bibnamefont{Kindratenko}},
  \bibinfo{author}{\bibfnamefont{G.~K.} \bibnamefont{Thiruvathukal}},
  \bibnamefont{and} \bibinfo{author}{\bibfnamefont{S.}~\bibnamefont{Gottlieb}},
  \bibinfo{journal}{Computing in Science \& Engineering}
  \textbf{\bibinfo{volume}{10}}, \bibinfo{pages}{13} (\bibinfo{year}{2008}),
  \urlprefix\url{http://scitation.aip.org/content/aip/journal/cise/10/6/10.1109/MCSE.2008.149}.

\bibitem[{\citenamefont{Schneider}(2015)}]{Schneider2015}
\bibinfo{author}{\bibfnamefont{B.~I.} \bibnamefont{Schneider}},
  \bibinfo{journal}{Computing in Science Engineering}
  \textbf{\bibinfo{volume}{17}}, \bibinfo{pages}{9} (\bibinfo{year}{2015}),
  ISSN \bibinfo{issn}{1521-9615}.

\bibitem[{\citenamefont{Ang et~al.}(2010)\citenamefont{Ang, Bergman, Borkar,
  Carlson, Carrington, Chiu, Colwell, Dally, Dongarra, Geist
  et~al.}}]{Lucas2014}
\bibinfo{author}{\bibfnamefont{J.}~\bibnamefont{Ang}},
  \bibinfo{author}{\bibfnamefont{K.}~\bibnamefont{Bergman}},
  \bibinfo{author}{\bibfnamefont{S.}~\bibnamefont{Borkar}},
  \bibinfo{author}{\bibfnamefont{W.}~\bibnamefont{Carlson}},
  \bibinfo{author}{\bibfnamefont{L.}~\bibnamefont{Carrington}},
  \bibinfo{author}{\bibfnamefont{G.}~\bibnamefont{Chiu}},
  \bibinfo{author}{\bibfnamefont{R.}~\bibnamefont{Colwell}},
  \bibinfo{author}{\bibfnamefont{W.}~\bibnamefont{Dally}},
  \bibinfo{author}{\bibfnamefont{J.}~\bibnamefont{Dongarra}},
  \bibinfo{author}{\bibfnamefont{A.}~\bibnamefont{Geist}},
  \bibnamefont{et~al.}, \bibinfo{type}{Tech. Rep.}, \bibinfo{institution}{DOE
  ASCAC Subcommittee Report} (\bibinfo{year}{2010}).

\bibitem[{\citenamefont{Ashby et~al.}(2010)\citenamefont{Ashby, Beckman, Chen,
  Colella, Collins, Crawford, Dongarra, Kothe, Lusk, Messina
  et~al.}}]{Ashby2010}
\bibinfo{author}{\bibfnamefont{S.}~\bibnamefont{Ashby}},
  \bibinfo{author}{\bibfnamefont{P.}~\bibnamefont{Beckman}},
  \bibinfo{author}{\bibfnamefont{J.}~\bibnamefont{Chen}},
  \bibinfo{author}{\bibfnamefont{P.}~\bibnamefont{Colella}},
  \bibinfo{author}{\bibfnamefont{B.}~\bibnamefont{Collins}},
  \bibinfo{author}{\bibfnamefont{D.}~\bibnamefont{Crawford}},
  \bibinfo{author}{\bibfnamefont{J.}~\bibnamefont{Dongarra}},
  \bibinfo{author}{\bibfnamefont{D.}~\bibnamefont{Kothe}},
  \bibinfo{author}{\bibfnamefont{R.}~\bibnamefont{Lusk}},
  \bibinfo{author}{\bibfnamefont{P.}~\bibnamefont{Messina}},
  \bibnamefont{et~al.}, \bibinfo{type}{Tech. Rep.},
  \bibinfo{institution}{Summary Report of the Advanced Scientific Computing
  Advisory Committee Subcommittee} (\bibinfo{year}{2010}).

\bibitem[{\citenamefont{Geist and Lucas}(2009)}]{Geist2009}
\bibinfo{author}{\bibfnamefont{A.}~\bibnamefont{Geist}} \bibnamefont{and}
  \bibinfo{author}{\bibfnamefont{R.}~\bibnamefont{Lucas}},
  \bibinfo{journal}{Int. J. of High. Perform. Comput. Appl.}
  \textbf{\bibinfo{volume}{23}}, \bibinfo{pages}{427–436}
  (\bibinfo{year}{2009}).

\bibitem[{\citenamefont{Simon}(1997)}]{Simon1997}
\bibinfo{author}{\bibfnamefont{D.~R.} \bibnamefont{Simon}},
  \bibinfo{journal}{SIAM Journal on Computing} \textbf{\bibinfo{volume}{26}},
  \bibinfo{pages}{1474} (\bibinfo{year}{1997}),
  \urlprefix\url{http://epubs.siam.org/doi/abs/10.1137/S0097539796298637}.

\bibitem[{\citenamefont{Shor}(1997)}]{Shor1997}
\bibinfo{author}{\bibfnamefont{P.~W.} \bibnamefont{Shor}},
  \bibinfo{journal}{SIAM Journal on Computing} \textbf{\bibinfo{volume}{26}},
  \bibinfo{pages}{1484} (\bibinfo{year}{1997}),
  \urlprefix\url{http://epubs.siam.org/doi/abs/10.1137/S0097539795293172}.

\bibitem[{\citenamefont{Kassal et~al.}(2011)\citenamefont{Kassal, Whitfield,
  Perdomo-Ortiz, Yung, and Aspuru-Guzik}}]{Kassal2010}
\bibinfo{author}{\bibfnamefont{I.}~\bibnamefont{Kassal}},
  \bibinfo{author}{\bibfnamefont{J.~D.} \bibnamefont{Whitfield}},
  \bibinfo{author}{\bibfnamefont{A.}~\bibnamefont{Perdomo-Ortiz}},
  \bibinfo{author}{\bibfnamefont{M.-H.} \bibnamefont{Yung}}, \bibnamefont{and}
  \bibinfo{author}{\bibfnamefont{A.}~\bibnamefont{Aspuru-Guzik}},
  \bibinfo{journal}{Annual Review of Physical Chemistry}
  \textbf{\bibinfo{volume}{62}}, \bibinfo{pages}{185} (\bibinfo{year}{2011}),
  \urlprefix\url{http://dx.doi.org/10.1146/annurev-physchem-032210-103512}.

\bibitem[{\citenamefont{Abrams and Lloyd}(1997)}]{Abrams1997}
\bibinfo{author}{\bibfnamefont{D.~S.} \bibnamefont{Abrams}} \bibnamefont{and}
  \bibinfo{author}{\bibfnamefont{S.}~\bibnamefont{Lloyd}},
  \bibinfo{journal}{Phys. Rev. Lett.} \textbf{\bibinfo{volume}{79}},
  \bibinfo{pages}{2586} (\bibinfo{year}{1997}),
  \urlprefix\url{http://link.aps.org/doi/10.1103/PhysRevLett.79.2586}.

\bibitem[{\citenamefont{van Nieuwpoort et~al.}(2001)\citenamefont{van
  Nieuwpoort, Kielmann, and Bal}}]{vanNieuwpoort2001}
\bibinfo{author}{\bibfnamefont{R.~V.} \bibnamefont{van Nieuwpoort}},
  \bibinfo{author}{\bibfnamefont{T.}~\bibnamefont{Kielmann}}, \bibnamefont{and}
  \bibinfo{author}{\bibfnamefont{H.~E.} \bibnamefont{Bal}},
  \bibinfo{journal}{SIGPLAN Not.} \textbf{\bibinfo{volume}{36}},
  \bibinfo{pages}{34 } (\bibinfo{year}{2001}), ISSN \bibinfo{issn}{0362-1340},
  \urlprefix\url{http://doi.acm.org/10.1145/568014.379563}.

\bibitem[{\citenamefont{Bauer et~al.}(2015)\citenamefont{Bauer, Wecker, Millis,
  Hastings, and Troyer}}]{Bauer2015}
\bibinfo{author}{\bibfnamefont{B.}~\bibnamefont{Bauer}},
  \bibinfo{author}{\bibfnamefont{D.}~\bibnamefont{Wecker}},
  \bibinfo{author}{\bibfnamefont{A.~J.} \bibnamefont{Millis}},
  \bibinfo{author}{\bibfnamefont{M.~B.} \bibnamefont{Hastings}},
  \bibnamefont{and} \bibinfo{author}{\bibfnamefont{M.}~\bibnamefont{Troyer}},
  \emph{\bibinfo{title}{Hybrid quantum-classical approach to correlated
  materials}} (\bibinfo{year}{2015}), \bibinfo{note}{arXiv:1510.03859
  [quant-ph]}.

\bibitem[{\citenamefont{Kreula et~al.}(2015)\citenamefont{Kreula, Clark, and
  Jaksch}}]{Kreula2015}
\bibinfo{author}{\bibfnamefont{J.~M.} \bibnamefont{Kreula}},
  \bibinfo{author}{\bibfnamefont{S.~R.} \bibnamefont{Clark}}, \bibnamefont{and}
  \bibinfo{author}{\bibfnamefont{D.}~\bibnamefont{Jaksch}},
  \emph{\bibinfo{title}{A quantum coprocessor for accelerating simulations of
  non-equilibrium many body quantum dynamics}} (\bibinfo{year}{2015}),
  \bibinfo{note}{arXiv:1510.05703 [quant-ph]}.

\bibitem[{\citenamefont{Nielsen and Chuang}(2000)}]{Nielsen2000}
\bibinfo{author}{\bibfnamefont{M.~A.} \bibnamefont{Nielsen}} \bibnamefont{and}
  \bibinfo{author}{\bibfnamefont{I.~L.} \bibnamefont{Chuang}},
  \emph{\bibinfo{title}{Quantum computation and quantum information}}
  (\bibinfo{publisher}{Cambridge University Press}, \bibinfo{year}{2000}).

\bibitem[{\citenamefont{{Deutsch}}(1985)}]{Deutsch1985}
\bibinfo{author}{\bibfnamefont{D.}~\bibnamefont{{Deutsch}}},
  \bibinfo{journal}{Proceedings of the Royal Society of London Series A}
  \textbf{\bibinfo{volume}{400}}, \bibinfo{pages}{97} (\bibinfo{year}{1985}).

\bibitem[{\citenamefont{Childs and van Dam}(2010)}]{Childs2010}
\bibinfo{author}{\bibfnamefont{A.~M.} \bibnamefont{Childs}} \bibnamefont{and}
  \bibinfo{author}{\bibfnamefont{W.}~\bibnamefont{van Dam}},
  \bibinfo{journal}{Rev. Mod. Phys.} \textbf{\bibinfo{volume}{82}},
  \bibinfo{pages}{1} (\bibinfo{year}{2010}),
  \urlprefix\url{http://link.aps.org/doi/10.1103/RevModPhys.82.1}.

\bibitem[{\citenamefont{Johnson et~al.}(2011)\citenamefont{Johnson, Amin,
  Gildert, Lanting, Hamze, Dickson, Harris, Berkley, Johansson, Bunyk
  et~al.}}]{Johnson2011}
\bibinfo{author}{\bibfnamefont{M.}~\bibnamefont{Johnson}},
  \bibinfo{author}{\bibfnamefont{M.}~\bibnamefont{Amin}},
  \bibinfo{author}{\bibfnamefont{S.}~\bibnamefont{Gildert}},
  \bibinfo{author}{\bibfnamefont{T.}~\bibnamefont{Lanting}},
  \bibinfo{author}{\bibfnamefont{F.}~\bibnamefont{Hamze}},
  \bibinfo{author}{\bibfnamefont{N.}~\bibnamefont{Dickson}},
  \bibinfo{author}{\bibfnamefont{R.}~\bibnamefont{Harris}},
  \bibinfo{author}{\bibfnamefont{A.}~\bibnamefont{Berkley}},
  \bibinfo{author}{\bibfnamefont{J.}~\bibnamefont{Johansson}},
  \bibinfo{author}{\bibfnamefont{P.}~\bibnamefont{Bunyk}},
  \bibnamefont{et~al.}, \bibinfo{journal}{Nature}
  \textbf{\bibinfo{volume}{473}}, \bibinfo{pages}{194} (\bibinfo{year}{2011}).

\bibitem[{\citenamefont{Saeedi et~al.}(2013)\citenamefont{Saeedi, Simmons,
  Salvail, Dluhy, Riemann, Abrosimov, Becker, Pohl, Morton, and
  Thewalt}}]{Saeedi2013}
\bibinfo{author}{\bibfnamefont{K.}~\bibnamefont{Saeedi}},
  \bibinfo{author}{\bibfnamefont{S.}~\bibnamefont{Simmons}},
  \bibinfo{author}{\bibfnamefont{J.~Z.} \bibnamefont{Salvail}},
  \bibinfo{author}{\bibfnamefont{P.}~\bibnamefont{Dluhy}},
  \bibinfo{author}{\bibfnamefont{H.}~\bibnamefont{Riemann}},
  \bibinfo{author}{\bibfnamefont{N.~V.} \bibnamefont{Abrosimov}},
  \bibinfo{author}{\bibfnamefont{P.}~\bibnamefont{Becker}},
  \bibinfo{author}{\bibfnamefont{H.-J.} \bibnamefont{Pohl}},
  \bibinfo{author}{\bibfnamefont{J.~J.~L.} \bibnamefont{Morton}},
  \bibnamefont{and} \bibinfo{author}{\bibfnamefont{M.~L.~W.}
  \bibnamefont{Thewalt}}, \bibinfo{journal}{Science}
  \textbf{\bibinfo{volume}{342}}, \bibinfo{pages}{830} (\bibinfo{year}{2013}),
  \urlprefix\url{http://www.sciencemag.org/content/342/6160/830.abstract}.

\bibitem[{\citenamefont{Monroe and Kim}(2013)}]{Monroe2013}
\bibinfo{author}{\bibfnamefont{C.}~\bibnamefont{Monroe}} \bibnamefont{and}
  \bibinfo{author}{\bibfnamefont{J.}~\bibnamefont{Kim}},
  \bibinfo{journal}{Science} \textbf{\bibinfo{volume}{339}},
  \bibinfo{pages}{1164} (\bibinfo{year}{2013}),
  \urlprefix\url{http://www.sciencemag.org/content/339/6124/1164.abstract}.

\bibitem[{\citenamefont{Devoret and Schoelkopf}(2013)}]{Devoret2013}
\bibinfo{author}{\bibfnamefont{M.~H.} \bibnamefont{Devoret}} \bibnamefont{and}
  \bibinfo{author}{\bibfnamefont{R.~J.} \bibnamefont{Schoelkopf}},
  \bibinfo{journal}{Science} \textbf{\bibinfo{volume}{339}},
  \bibinfo{pages}{1169} (\bibinfo{year}{2013}),
  \urlprefix\url{http://www.sciencemag.org/content/339/6124/1169.abstract}.

\bibitem[{\citenamefont{Meter and Oskin}(2006)}]{vanMeter2006}
\bibinfo{author}{\bibfnamefont{R.~v.} \bibnamefont{Meter}} \bibnamefont{and}
  \bibinfo{author}{\bibfnamefont{M.}~\bibnamefont{Oskin}},
  \bibinfo{journal}{ACM Journal on Emerging Technologies in Computing Systems
  (JETC)} \textbf{\bibinfo{volume}{2}}, \bibinfo{pages}{31}
  (\bibinfo{year}{2006}).

\bibitem[{\citenamefont{Thaker et~al.}(2006)\citenamefont{Thaker, Metodi,
  Cross, Chuang, and Chong}}]{Thaker2006}
\bibinfo{author}{\bibfnamefont{D.~D.} \bibnamefont{Thaker}},
  \bibinfo{author}{\bibfnamefont{T.~S.} \bibnamefont{Metodi}},
  \bibinfo{author}{\bibfnamefont{A.~W.} \bibnamefont{Cross}},
  \bibinfo{author}{\bibfnamefont{I.~L.} \bibnamefont{Chuang}},
  \bibnamefont{and} \bibinfo{author}{\bibfnamefont{F.~T.} \bibnamefont{Chong}},
  \bibinfo{journal}{SIGARCH Comput. Archit. News}
  \textbf{\bibinfo{volume}{34}}, \bibinfo{pages}{378} (\bibinfo{year}{2006}),
  ISSN \bibinfo{issn}{0163-5964},
  \urlprefix\url{http://doi.acm.org/10.1145/1150019.1136518}.

\bibitem[{\citenamefont{Van~Meter et~al.}(2010)\citenamefont{Van~Meter, Ladd,
  Fowler, and Yamamoto}}]{vanMeter2010}
\bibinfo{author}{\bibfnamefont{R.}~\bibnamefont{Van~Meter}},
  \bibinfo{author}{\bibfnamefont{T.~D.} \bibnamefont{Ladd}},
  \bibinfo{author}{\bibfnamefont{A.~G.} \bibnamefont{Fowler}},
  \bibnamefont{and} \bibinfo{author}{\bibfnamefont{Y.}~\bibnamefont{Yamamoto}},
  \bibinfo{journal}{International Journal of Quantum Information}
  \textbf{\bibinfo{volume}{8}}, \bibinfo{pages}{295} (\bibinfo{year}{2010}).

\bibitem[{\citenamefont{Jones et~al.}(2012)\citenamefont{Jones, Van~Meter,
  Fowler, McMahon, Kim, Ladd, and Yamamoto}}]{Jones2012}
\bibinfo{author}{\bibfnamefont{N.~C.} \bibnamefont{Jones}},
  \bibinfo{author}{\bibfnamefont{R.}~\bibnamefont{Van~Meter}},
  \bibinfo{author}{\bibfnamefont{A.~G.} \bibnamefont{Fowler}},
  \bibinfo{author}{\bibfnamefont{P.~L.} \bibnamefont{McMahon}},
  \bibinfo{author}{\bibfnamefont{J.}~\bibnamefont{Kim}},
  \bibinfo{author}{\bibfnamefont{T.~D.} \bibnamefont{Ladd}}, \bibnamefont{and}
  \bibinfo{author}{\bibfnamefont{Y.}~\bibnamefont{Yamamoto}},
  \bibinfo{journal}{Physical Review X} \textbf{\bibinfo{volume}{2}},
  \bibinfo{pages}{031007} (\bibinfo{year}{2012}).

\bibitem[{\citenamefont{Metodi et~al.}(2011)\citenamefont{Metodi, Faruque, and
  Chong}}]{Metodi2011}
\bibinfo{author}{\bibfnamefont{T.~S.} \bibnamefont{Metodi}},
  \bibinfo{author}{\bibfnamefont{A.~I.} \bibnamefont{Faruque}},
  \bibnamefont{and} \bibinfo{author}{\bibfnamefont{F.~T.} \bibnamefont{Chong}},
  \emph{\bibinfo{title}{Quantum Computing for Computer Architects, Second
  Edition}}, Synthesis Lectures on Computer Architecture
  (\bibinfo{publisher}{Morgan {\&} Claypool Publishers}, \bibinfo{year}{2011}),
  \urlprefix\url{http://dx.doi.org/10.2200/S00331ED1V01Y201101CAC013}.

\bibitem[{\citenamefont{Hill et~al.}(2015)\citenamefont{Hill, Peretz, Hile,
  House, Fuechsle, Rogge, Simmons, and Hollenberg}}]{Hill2015}
\bibinfo{author}{\bibfnamefont{C.~D.} \bibnamefont{Hill}},
  \bibinfo{author}{\bibfnamefont{E.}~\bibnamefont{Peretz}},
  \bibinfo{author}{\bibfnamefont{S.~J.} \bibnamefont{Hile}},
  \bibinfo{author}{\bibfnamefont{M.~G.} \bibnamefont{House}},
  \bibinfo{author}{\bibfnamefont{M.}~\bibnamefont{Fuechsle}},
  \bibinfo{author}{\bibfnamefont{S.}~\bibnamefont{Rogge}},
  \bibinfo{author}{\bibfnamefont{M.~Y.} \bibnamefont{Simmons}},
  \bibnamefont{and} \bibinfo{author}{\bibfnamefont{L.~C.~L.}
  \bibnamefont{Hollenberg}}, \bibinfo{journal}{Science Advances}
  \textbf{\bibinfo{volume}{1}} (\bibinfo{year}{2015}).

\bibitem[{\citenamefont{Knill}(1996)}]{Knill1996}
\bibinfo{author}{\bibfnamefont{E.}~\bibnamefont{Knill}}, \bibinfo{type}{Tech.
  Rep.}, \bibinfo{institution}{Technical Report LAUR-96-2724, Los Alamos
  National Laboratory} (\bibinfo{year}{1996}).

\bibitem[{\citenamefont{Selinger}(2004)}]{Selinger2004}
\bibinfo{author}{\bibfnamefont{P.}~\bibnamefont{Selinger}},
  \bibinfo{journal}{Mathematical Structures in Computer Science}
  \textbf{\bibinfo{volume}{14}}, \bibinfo{pages}{527} (\bibinfo{year}{2004}),
  ISSN \bibinfo{issn}{1469-8072},
  \urlprefix\url{http://journals.cambridge.org/article_S0960129504004256}.

\bibitem[{\citenamefont{Green et~al.}(2013)\citenamefont{Green, Lumsdaine,
  Ross, Selinger, and Valiron}}]{Green2013}
\bibinfo{author}{\bibfnamefont{A.~S.} \bibnamefont{Green}},
  \bibinfo{author}{\bibfnamefont{P.~L.} \bibnamefont{Lumsdaine}},
  \bibinfo{author}{\bibfnamefont{N.~J.} \bibnamefont{Ross}},
  \bibinfo{author}{\bibfnamefont{P.}~\bibnamefont{Selinger}}, \bibnamefont{and}
  \bibinfo{author}{\bibfnamefont{B.}~\bibnamefont{Valiron}}, in
  \emph{\bibinfo{booktitle}{Proceedings of the 34th ACM SIGPLAN Conference on
  Programming Language Design and Implementation}} (\bibinfo{publisher}{ACM},
  \bibinfo{address}{New York, NY, USA}, \bibinfo{year}{2013}), PLDI '13, pp.
  \bibinfo{pages}{333--342}, ISBN \bibinfo{isbn}{978-1-4503-2014-6},
  \urlprefix\url{http://doi.acm.org/10.1145/2491956.2462177}.

\bibitem[{\citenamefont{Wecker and Svore}(2014)}]{Wecker2014}
\bibinfo{author}{\bibfnamefont{D.}~\bibnamefont{Wecker}} \bibnamefont{and}
  \bibinfo{author}{\bibfnamefont{K.~M.} \bibnamefont{Svore}},
  \emph{\bibinfo{title}{LIQUID: A Software Design Architecture and
  Domain-Specific Language for Quantum Computing}} (\bibinfo{year}{2014}),
  \bibinfo{note}{\url{http://arxiv.org/pdf/1402.4467v1.pdf}}.

\bibitem[{\citenamefont{Abhari et~al.}(2012)\citenamefont{Abhari, Faruque,
  Dousti, Svec, Catu, Chakrabati, Chiang, Vanderwilt, Black, Chong
  et~al.}}]{Abhari2012}
\bibinfo{author}{\bibfnamefont{A.~J.} \bibnamefont{Abhari}},
  \bibinfo{author}{\bibfnamefont{A.}~\bibnamefont{Faruque}},
  \bibinfo{author}{\bibfnamefont{M.~J.} \bibnamefont{Dousti}},
  \bibinfo{author}{\bibfnamefont{L.}~\bibnamefont{Svec}},
  \bibinfo{author}{\bibfnamefont{O.}~\bibnamefont{Catu}},
  \bibinfo{author}{\bibfnamefont{A.}~\bibnamefont{Chakrabati}},
  \bibinfo{author}{\bibfnamefont{C.-F.} \bibnamefont{Chiang}},
  \bibinfo{author}{\bibfnamefont{S.}~\bibnamefont{Vanderwilt}},
  \bibinfo{author}{\bibfnamefont{J.}~\bibnamefont{Black}},
  \bibinfo{author}{\bibfnamefont{F.}~\bibnamefont{Chong}},
  \bibnamefont{et~al.}, \bibinfo{type}{Tech. Rep.} (\bibinfo{year}{2012}),
  \urlprefix\url{ftp://ftp.cs.princeton.edu/techreports/2012/934.pdf}.

\bibitem[{\citenamefont{Hornibrook et~al.}(2015)\citenamefont{Hornibrook,
  Colless, Conway~Lamb, Pauka, Lu, Gossard, Watson, Gardner, Fallahi, Manfra
  et~al.}}]{Hornibrook2015}
\bibinfo{author}{\bibfnamefont{J.~M.} \bibnamefont{Hornibrook}},
  \bibinfo{author}{\bibfnamefont{J.~I.} \bibnamefont{Colless}},
  \bibinfo{author}{\bibfnamefont{I.~D.} \bibnamefont{Conway~Lamb}},
  \bibinfo{author}{\bibfnamefont{S.~J.} \bibnamefont{Pauka}},
  \bibinfo{author}{\bibfnamefont{H.}~\bibnamefont{Lu}},
  \bibinfo{author}{\bibfnamefont{A.~C.} \bibnamefont{Gossard}},
  \bibinfo{author}{\bibfnamefont{J.~D.} \bibnamefont{Watson}},
  \bibinfo{author}{\bibfnamefont{G.~C.} \bibnamefont{Gardner}},
  \bibinfo{author}{\bibfnamefont{S.}~\bibnamefont{Fallahi}},
  \bibinfo{author}{\bibfnamefont{M.~J.} \bibnamefont{Manfra}},
  \bibnamefont{et~al.}, \bibinfo{journal}{Phys. Rev. Applied}
  \textbf{\bibinfo{volume}{3}}, \bibinfo{pages}{024010} (\bibinfo{year}{2015}),
  \urlprefix\url{http://link.aps.org/doi/10.1103/PhysRevApplied.3.024010}.

\bibitem[{\citenamefont{Carolan et~al.}(2015)\citenamefont{Carolan, Harrold,
  Sparrow, Martín-López, Russell, Silverstone, Shadbolt, Matsuda, Oguma, Itoh
  et~al.}}]{Carolan2015}
\bibinfo{author}{\bibfnamefont{J.}~\bibnamefont{Carolan}},
  \bibinfo{author}{\bibfnamefont{C.}~\bibnamefont{Harrold}},
  \bibinfo{author}{\bibfnamefont{C.}~\bibnamefont{Sparrow}},
  \bibinfo{author}{\bibfnamefont{E.}~\bibnamefont{Martín-López}},
  \bibinfo{author}{\bibfnamefont{N.~J.} \bibnamefont{Russell}},
  \bibinfo{author}{\bibfnamefont{J.~W.} \bibnamefont{Silverstone}},
  \bibinfo{author}{\bibfnamefont{P.~J.} \bibnamefont{Shadbolt}},
  \bibinfo{author}{\bibfnamefont{N.}~\bibnamefont{Matsuda}},
  \bibinfo{author}{\bibfnamefont{M.}~\bibnamefont{Oguma}},
  \bibinfo{author}{\bibfnamefont{M.}~\bibnamefont{Itoh}}, \bibnamefont{et~al.},
  \bibinfo{journal}{Science} \textbf{\bibinfo{volume}{349}},
  \bibinfo{pages}{711} (\bibinfo{year}{2015}),
  \urlprefix\url{http://www.sciencemag.org/content/349/6249/711.abstract}.

\bibitem[{\citenamefont{Broadbent et~al.}(2009)\citenamefont{Broadbent,
  Fitzsimons, and Kashefi}}]{Broadbent2009}
\bibinfo{author}{\bibfnamefont{A.}~\bibnamefont{Broadbent}},
  \bibinfo{author}{\bibfnamefont{J.}~\bibnamefont{Fitzsimons}},
  \bibnamefont{and} \bibinfo{author}{\bibfnamefont{E.}~\bibnamefont{Kashefi}}
  (\bibinfo{year}{2009}).

\end{thebibliography}

\end{document}